# Резко асимметричная дифракция как метод определения магнитооптических констант для рентгеновского излучения вблизи краев поглощения

*М. А. Андреева*[1*], *Ю. Л. Репченко*[2**], *А.Г.Смехова*[1],
*К. Думенил*[3], *Ф. Вилхелм*[4], *А. Рогалев*[4]

[1] *Физический факультет, МГУ имени М.В. Ломоносова, 119991, Ленинские горы 1, стр.2, Москва*

[2] *Физический факультет, Воронежский Государственный Университет, 394036, Университетская пл. 1, Воронеж, Россия*

[3] *Институт Жана Ламура, Университет Лотарингии,* 54011 *Нанси, Франция* (Institut *Jean Lamour (UMR CNRS 7198), University of Lorraine, 54011 Vandoeuvre-les-Nancy Cedex, France)*

[4] *Европейский источник синхротронного излучения (ESRF), Гренобль, BP 220, 38043, Франция (European Synchrotron Radiation Facility, BP 220, 38043 Grenoble Cedex, France).*

*E-mail: Mandreeva1@yandex.ru; **E-mail: kent160@mail.ru

Аннотация.

Проанализирована спектральная зависимость положения брэгговского максимума в условиях резко асимметричной дифракции в рамках кинематического и динамического приближения теории дифракции. Модельные расчеты выполнены для $L_3$ края поглощения Y в монокристаллической плёнке $YFe_2$, они продемонстрировали возможность определения по этой зависимости магнитооптических констант (или, эквивалентно, дисперсионных поправок к атомному фактору рассеяния) для жесткого рентгеновского излучения. Проведено сравнение с экспериментальными данными, полученными для образца Nb(4 нм)/$YFe_2$(40 нм <110>)/Fe(1.5 нм)/Nb(50 нм)/сапфир на Европейском источнике синхротронного излучения.

## 1. ВВЕДЕНИЕ

Магнитооптические константы вблизи краев поглощения рентгеновского излучения и напрямую связанные с ними дисперсионные поправки к атомному фактору рассеяния важно знать для проведения элементно- и пространственно-селективных исследования электронной и магнитной структуры атомов в мультислоях, кристаллах и сложных магнитных структурах методами XRMR (X-ray Resonant Magnetic Reflectometry) и DAFS (Diffraction Anomalous Fine Structure), для интерпретации волноводных и других эффектов, наблюдаемых вблизи краев поглощения (см., например, [1-8]). Отметим, что эти направления исследований бурно развиваются на всех синхротронах в мире, предоставляя уникальную информацию, углубляя наши фундаментальные представления об электронной структуре конденсированных сред и природе магнетизма.

Для мягкого рентгеновского излучения преломление и поглощение, включая магнитные поправки, (то есть $\delta$ и $\beta$ в показателе преломления $n = 1 + \delta + i\beta$) могут быть экспериментально определены из сдвига брэгговских максимумов на кривых отражения от периодических мультислоев [9-13]. Для жёсткого рентгеновского излучения брэгговские максимумы возникают при дифракции на кристаллах, но влияние преломления на их сдвиг слишком мало, чтобы его можно было использовать для определения дисперсионных поправок к показателю преломления и, соответственно, атомному фактору рассеяния. Преломление для жесткого рентгеновского излучения существенно лишь вблизи критического угла полного внешнего отражения, а брэгговское отражение от кристаллической структуры в симметричной геометрии имеет место при углах скольжения существенно больших критического. Поэтому именно рефлектометрия (зеркальное отражение при скользящих углах) наиболее успешно применяется для определения $\delta$ и $\beta$ (см., например, [14]), в том числе и для определения магнитных добавок к показателю преломления [15]. Однако, этот метод предъявляет высокие требования к поверхности исследуемых образцов и требует тщательного приготовления пленок, которое подразумевает включение буферных и защитных слоев. При этом интерпретация рефлектометрических кривых от многослойного объекта становится достаточно сложной и неоднозначной.

Можно ожидать, что влияние преломления на сдвиг брэгговских максимумов от кристаллической структуры (и, соответственно, возможность определения дисперсионных поправок) возрастает в резко асимметричной схеме дифракции при скользящих углах падения, что позволит использовать этот более простой метод для определения магнитооптических констант. Такую возможность мы анализируем в предлагаемой статье, используя как стартовый момент обнаружение неожиданных эффектов в экспериментальных данных по рефлектометрии.

## 2. ЭКСПЕРИМЕНТ

В недавно проведенном эксперименте на ESRF по определению спектральных зависимостей магнитооптических констант вблизи $L_{2,3}$ краев поглощения иттрия на образце Nb(4 *нм*)/$YFe_2$(40 *нм* <110>)/Fe(1.5 *нм*)/Nb(50 *нм*)/сапфир [15] измерялись рефлектометрические кривые (зависимость отражения от угла скольжения) для для набора энергий вблизи $L_{2,3}$ краев поглощения Y. Аппаратура была настроена так, что одновременно с зеркально отраженным сигналом регистрировалось и флуоресцентное излучение от этого образца в функции угла скольжения (Рис. 1а-б). Флуоресцентный детектор располагался над поверхностью образца. Эти данные не входили в основную задачу: определения магнитных добавок к тензору восприимчивости по сдвигу осцилляций Кизиха на рефлектометрических кривых, - и не представляли на первый взгляд никакого интереса. Флуоресцентный детектор не имел разрешения по энергии, так что регистрировалось суммарное флуоресцентное излучение от всех атомов в исследуемом образце: $L_{\alpha12}$ Y (1920.47 эВ), $L_{\beta1}$ Y (1995.84 эВ), $K_{\alpha1,2}$ Al (1486,27 эВ), $K_{\beta1}$ Al (1557,45 эВ), $L_{\alpha1,2}$ Fe (705 эВ), $L_{\beta1}$ Fe (718.5 эВ), и со слабой эффективностью $K_{\alpha1}$ O (524,9 эВ), - для энергии падающих фотонов в интервале 2071 – 2095 эВ. Для такого смешанного сигнала корректная обработка затруднительно, но в целом поведение флуоресцентных кривых вполне согласуется с предположением, что основной сигнал поступает из пленки $YFe_2$ и подложки $Al_2O_3$ за исключением одной маленькой детали. На всех флуоресцентных кривых (Рис. 1а-б) имеется пичок в области малых углов скольжения, который можно было бы и проигнорировать как дефект измерительной аппаратуры, но его положение изменяется с энергией падающих фотонов вполне систематически и даже как-то (хотя и слабо) согласуется с резонансной зависимостью восприимчивости в исследуемой области (Рис. 2). Была предпринята попытка (неудавшаяся) описать такую зависимость изменением критического угла полного внешнего отражения при варьировании энергии фотонов от какой-либо границы раздела. Хорошо известно, что вблизи критического угла возникает всплеск вторичного излучения [16-18]. Проанализировали также возможность формирования волноводного режима в пленке $YFe_2$ [19], при котором возникают локальные (по углу и по пространству) усиления возбуждающего поля в образце и соответствующие пики в выходе вторичного излучения, как это иллюстрируется Рисунком 3. Параметры структуры для расчета были выбраны идеальными для формирования стоячей волны (увеличена оптическая плотность Nb, то есть $\mathrm{Re}\chi_{Nb}$, и уменьшено поглощение в слое $YFe_2$). Из Рис. 3 видно, что пучности поля возникают для углов, существенно бо́льших критического угла полного отражения, что противоречит нашим экспериментальным данным. Кроме того, идеальные параметры волноводной структуры не соответствовали оптическим характеристикам образца, полученным в [15]. Самым существенным опровержением волноводного объяснения пичков явилась слабая зависимость смещения пичков от энергии фотонов в этой модели. Близкая к линейной экспериментальная зависимость положения пичков от энергии падающих фотонов, интерпретируемая в простейшем приближении $2d\sin\theta = n\lambda$, дала для дифракционной толщины $d \approx 0.4$ нм. Это обстоятельство привело к мысли, что кроме флуоресцентного излучения в детектор случайно попал сигнал дифракции на кристалле $YFe_2$. Используя пакет PowderCell [20], нашли, что для наших очень больших длин волн ($\lambda$=0.591 – 0.598 нм) все же существует единственное дифракционное отражение (111) от структуры C15 ($YFe_2$) с параметром решетки $d$=0.4255 нм. Обнаруженный эффект простимулировал рассмотрение влияния преломления на положение дифракционного пика в условиях резко асимметричной дифракции в скользящей геометрии.

## 3. ТЕОРИЯ

В простейшем случае без учета преломления положение брэгговского максимума определяется законом Вульфа-Брэгга

$$2d\sin(\theta + \psi) = n\lambda, \qquad (1)$$

где $\theta$ угол скольжения для падающего излучения, $\psi$ - угол между поверхностью кристалла и атомно-кристаллическими плоскостями (Рис. 4), d – межплоскостное расстояние, n – порядок отражения. При скользящих углах падения $\sin(\theta + \psi) \approx (\theta\cos\psi + \sin\psi)$, что дает линейную зависимость положения угла Брэгга от длины волны (или обратно пропорциональную от энергии фотона $E_{ph}$).

В случае кинематической дифракции с учетом преломления волновой вектор в среде равен $\mathbf{k}_1 = \frac{2\pi}{\lambda}\{0, \cos\theta, \sin\theta'\}$, $|\mathbf{k}_1| = \frac{2\pi}{\lambda}(1 + \chi/2)$, где

$$\sin\theta' = \eta = \sqrt{\sin^2\theta + \chi}\,, \tag{2}$$

$\eta$- нормальная компонента волнового вектора в единицах $2\pi/\lambda = \omega/c$,

$$\chi = 2\delta + 2i\beta = \frac{\lambda^2}{\pi}\rho f \tag{3}$$

- восприимчивость среды, $\rho$ - объемная плотность атомов, f - атомный фактор рассеяния на нулевой угол. В (3) представлена известная связь между восприимчивостью, показателем преломления и атомным фактором рассеяния для рентгеновского излучения, отмеченная в начале статьи.

Для набега фазы при отражении от кристаллических плоскостей на одном периоде должно быть

$$\exp(-2i\,\mathrm{Re}\,k_{1h}d) = \exp(2\pi i), \tag{4}$$

где $k_{1h}$ - проекция (комплексного) волнового вектора преломленной волны на нормаль к отражающим плоскостям (то есть на направление вектора обратной решетки $\mathbf{h} = \frac{2\pi}{d}\{0,-\sin\psi,-\cos\psi\}$). Косинус угла между $\mathbf{h}$ и $\mathbf{k}_1$ (или минус синус угла преломленной волны с отражающей кристаллической плоскостью) равен

$$\frac{(\mathbf{k}_1\mathbf{h})}{|k_1||h|} = \cos\alpha = \frac{-\cos\theta\sin\psi - \mathrm{Re}(\sin\theta')\cos\psi}{1+\chi/2}. \tag{5}$$

Таким образом

$$-\mathrm{Re}\,k_{1h} = \frac{2\pi}{\lambda}(\cos\theta\sin\psi + \mathrm{Re}(\sin\theta')\cos\psi), \tag{6}$$

и мы получаем обобщение формулы Вульфа-Брэгга на случай резко асимметричной дифракции в виде:

$$\sin\psi\cos\theta_m + \cos\psi\,\mathrm{Re}\sqrt{\sin^2\theta_m + \chi(E_{ph})} = \frac{\lambda}{2d}. \tag{7}$$

Это уравнение для определения угла скольжения $\theta_m(E_{ph})$, при котором возникает брэгговское отражение, мы решаем методом

Ньютона, не используя процедуру приближенного вычисления квадратного корня. Отметим, что в работах [9-13, 21] квадратный корень в (7) вычислялся приближенно (что справедливо только вдали от критического угла, но это не работает в нашем случае), поэтому считалось, что в кинематическом приближении мнимая часть $\chi(E_{ph})$ не влияет на положение брэгговского максимума. Точное решение (7) учитывает $\mathrm{Im}\chi(E_{ph})$, однако влияние поглощения в этом алгоритме все же невелико.

Мнимая часть фазового сдвига $2\mathrm{Im}k_{1h}d$ искажает форму брэгговского пика, и если поглощение асимметрично относительно $\theta_m$, максимум сдвигается. В кинематическом приближении форму и, соответственно, точное положение брэгговского максимума дает функция Лауэ (как сумма бесконечной геометрической прогрессии для волн, рассеянных кристаллическими плоскостями [22]). В нашем случае мы представляем ее в виде:

$$I \sim \left| t \frac{1}{1 - e^{-2ik_{1h}d}} \right|^2, \tag{8}$$

где $t$ – амплитуда преломленной волны, которая для случая границы между внешней средой и полубесконечным кристаллом определяется формулой Френеля $t = \dfrac{2\sin\theta}{\sin\theta + \eta}$.

Динамическая теория резко асимметричной дифракции рассматривалась ранее в основном для решения задач управления пучками рентгеновского излучения и были получены приближенные выражения для ширины, сдвига и расходимости дифрагированного пучка [23-26]. Эффект усиления влияния преломления на эти параметры был продемонстрирован и даже использовался для разделения близких дифракционных максимумов от пленки и подложки [27]. Для нашей задачи мы будем решать дисперсионное уравнение динамической теории 4-ой степени, не пренебрегая квадратичными по преломлению добавками к квадратам волновых векторов, и находить положение брэгговского пика $\theta_m(E_{ph})$ как экстремумом модуля в квадрате амплитуды дифрагированной волны.

В динамической теории с учетом пространственной периодичности кристалла диэлектрическую восприимчивость $\chi(\mathbf{r})$ представляют рядом Фурье

$$\chi(\mathbf{r}) = \sum_{h} \chi_h e^{-i\frac{\omega}{c}\boldsymbol{\tau}\mathbf{r}}, \tag{9}$$

что в двухволновом случае дает для амплитуд плоских волн проходящей $\mathbf{E}_1$ и дифрагированной $\mathbf{E}_2$ волн уравнения связи вида:

$$\begin{aligned} \mathbf{D}_1 &= (1+\chi_0)\mathbf{E}_1 + \chi_h \mathbf{E}_2 \\ \mathbf{D}_2 &= (1+\chi_0)\mathbf{E}_2 + \chi_{\bar{h}} \mathbf{E}_1 \end{aligned}. \tag{10}$$

Уравнения Максвелла для плоских волн $\sim\exp(i\frac{\omega}{c}\hat{\boldsymbol{\kappa}}_i\,\mathbf{r} - i\omega t)$ имеют вид:

$$\mathbf{D}_i = -[\hat{\boldsymbol{\kappa}}_i \mathbf{H}_i], \quad \mathbf{H}_i = [\hat{\boldsymbol{\kappa}}_i \mathbf{E}_i] \tag{11}$$

(в (9), (11) и далее используем волновые векторы и вектор обратной решетки $\boldsymbol{\tau}$ в единицах $2\pi/\lambda = \omega/c$, то есть $\boldsymbol{\tau} = \lambda/2\pi\,\mathbf{h}$, $\hat{\mathbf{k}}_2 = \hat{\mathbf{k}}_1 + \boldsymbol{\tau}$). Из (10), (11) получаем систему уравнений относительно амплитуд поля $\mathbf{E}_1$ и $\mathbf{E}_2$, которая в простейшем случае для $\sigma$-поляризации, когда все векторы $\mathbf{E}_i$ перпендикулярны плоскости рассеяния и мы можем работать со скалярными амплитудами, имеет вид:

$$\begin{aligned} (1+\chi_0)E_1 + \chi_h E_2 &= \hat{\boldsymbol{\kappa}}_1{}^2 E_1 \\ (1+\chi_0)E_2 + \chi_{\bar{h}} E_1 &= \hat{\boldsymbol{\kappa}}_2{}^2 E_2 \end{aligned} \tag{12}$$

(В общем случае векторы $\mathbf{E}_i$ не поперечны волновым векторам $\hat{\boldsymbol{\kappa}}_i$, поэтому для других поляризаций и анизотропных сред удобнее писать систему динамических уравнений для поперечных векторов $\mathbf{D}_i$ или

$\mathbf{H}_i$ [28, 29]). Система (13) имеет решение при условии:

$$(1-\hat{\mathbf{k}}_1^2+\chi_0)(1-\hat{\mathbf{k}}_2^2+\chi_0)-\chi_{\bar{h}}\chi_h=0. \qquad (13)$$

Это дисперсионное уравнение является уравнением 4-ой степени относительно поправки на преломление $\xi=\eta-\sin\theta$, так как

$$\hat{\mathbf{k}}_1^2=(\hat{\mathbf{k}}_0+\xi\mathbf{q})^2=1+2\xi\sin\theta+\xi^2, \qquad (14)$$

$$\begin{aligned}\hat{\mathbf{k}}_2^2&=(\hat{\mathbf{k}}_1+\boldsymbol{\tau})^2=\\&=(\hat{\mathbf{k}}_0+\xi\mathbf{q})^2+\mathbf{2\tau(}\hat{\mathbf{k}}_0+\xi\mathbf{q})+\boldsymbol{\tau}^2)=1+2\xi\sin\theta_h+\xi^2+\delta\end{aligned}, \qquad (15)$$

где $\hat{\mathbf{k}}_0$ - волновой вектор падающей волны единицах $\omega/c$ ($|\hat{\mathbf{k}}_0|^2=1$, $\mathbf{q}$ - единичный вектор нормали к поверхности, $\sin\theta_h=(\sin\theta+\boldsymbol{\tau}\mathbf{q})$ - угол выхода дифрагированной волны, $\delta=\boldsymbol{\tau}(\boldsymbol{\tau}+\mathbf{2}\hat{\mathbf{k}}_0)$ - отклонение от точного угла Брэгга. Если в (16) пренебречь $\xi^2$ (поскольку $\sin\theta_h$ достаточно велик) можно с достаточной точностью ограничиться решением уравнения (13) третьей степени по $\xi$, но это не слишком упрощает компьютерное решение задачи.

Граничная задача для определения амплитуд зеркально отраженной $E_{\mathbf{R}}$ от полубесконечного кристалла и дифрагированной волн $E_{\mathbf{h}}$ связывает тангенциальные компоненты электрического и магнитного поля излучения внутри и вне кристалла, что в нашем случае приводит к системе уравнений:

$$\begin{cases}E_0+E_R=\sum\limits_i E_1^{(i)}\\ \sin\theta(E_0-E_R)=\sum\limits_i(\sin\theta+\xi^{(i)})E_1^{(i)}\\ E_h=\sum\limits_i E_2^{(i)}\\ -\sin\theta_h E_h=-\sum\limits_i(\sin\theta_h+\xi^{(i)})E_2^{(i)}\end{cases} \qquad (16)$$

Для решения (16) следует учесть, что в кристалле амплитуды дифрагированных волн $E_2^{(i)}$ связаны с амплитудами проходящих волн $E_1^{(i)}$ соотношением, следующим из (12):

$$E_2^{(i)} = -\frac{1-\hat{k}_1^{(i)2}+\chi_0}{\chi_h}E_1^{(i)} = -\frac{\chi_{\bar{h}}}{1-\hat{k}_2^{(i)2}+\chi_0}E_1^{(i)}, \qquad (17)$$

Кроме того, анализ корней показывает, что в нашем случае резко ассиметричной дифракции для полубесконечной среды только один корень из четырех имеет физический смысл. При этом два первых уравнения в системе (16) становятся независимыми от двух последующих, и они фактически дают формулу Френеля для зеркально отражённой $E_R$ и единственной преломлённой $E_1^{(1)}$ волны (но в динамическом случае нормальная компонента волнового вектора в среде $\eta = \sin\theta + \xi^{(1)}$ определяется из уравнения (13), а не формулой (2)). Для дифрагированной волны $E_2^{(1)}$ преломление не существенно, поскольку волновой вектор дифрагированной волны образует достаточно большой угол с поверхностью (Рис. 4), так что можно вместо двух последних уравнений в системе (16) положить $E_h = E_2^{(1)}$. С учетом (17) это полностью решает задачу вычисления амплитуды дифрагированной волны как функцию угла скольжения падающего излучения $\theta$ и нахождения ее экстремума.

## 4. МОДЕЛИРОВАНИЕ

Численное моделирование мы провели для плёнки $YFe_2$ вблизи $L_3$ края поглощения Y ($E_{ph}$=2071 - 2095 эВ , $\lambda$=0.5986 – 0.5918 нм). Для отражения от кристаллических плоскостей (111) с межплоскостным расстоянием d=0.4255 нм для рассматриваемого интервала длин волн идеальный угол Брэгга составляет 44.71° – 44.06°. Для того, чтобы дифракция возникала при скользящем угле падения, мы выбрали угол $\psi$=43.9°. Спектральные зависимости реальной и мнимой части восприимчивости $\chi_0(E_{ph})$ мы взяли из работы [15]. Фурье компоненты $\chi_{h,\bar{h}}(E_{ph})$ вычислялись для структуры C15 ($Fd3mO_h^7$, атомы Y в положении 8a, атомы Fe в положении 16d) с использованием [30].

Уравнение (13) решалось с помощью программного пакета Wolfram Mathematica, который дал аналитическое решение уравнения 4-ой степени (слишком громоздкое, чтобы его здесь привести), и далее численный результат по вводимым параметрам. Для энергии $E_{ph}$=2082 эВ результат приведен на Рис. 5. Наглядно показано, что только один корень $\hat{\mathbf{k}}_{\mathbf{1}}^{(1)}$ в нашей геометрии имеет смысл. Волновой вектор $\hat{\mathbf{k}}_{\mathbf{1}}^{(2)}$ имеет небольшую отрицательную мнимую часть при положительной реальной части нормальной составляющей и не имеет физического смысла; для $\hat{\mathbf{k}}_{\mathbf{1}}^{(3)}$ волна направлена из среды, эта волна возбуждается в пленке при наличии нижней границы, $\hat{\mathbf{k}}_{\mathbf{1}}^{(4)}$ имеет слишком большую реальную часть, соответствующая дифрагированная волна согласно (17) пренебрежимо мала. Отметим, что $\eta^{(1)}$ очень мало отличается от $\eta = \sqrt{\sin^2\theta + \chi}$, что может быть обусловлено относительно малой величиной $\chi_h\chi_{\bar{h}}(E_{ph})$ для нашей структуры.

Для сравнения кинематики и динамики мы рассчитали форму дифракционного максимума $|E_h(\theta)|^2$ по формуле (8) и по динамической теории - (Рис. 6). (Отметим, что вычисленный «динамический» дифракционный максимум полностью совпадает с расчетом по формулам из работы [23].) Видно, что форма дифракционного максимума в кинематическом и динамическом приближении различается, но положение брэгговского максимума $\theta_m$ в обеих приближениях почти совпадает. В нашей задаче интересно именно положение брэгговского максимума $\theta_m$ в функции энергии фотонов $E_{ph}$, представленное на Рис. 7-8. Эти расчетные зависимости демонстрируют существенно более сильную зависимость положения брэгговского пика от эффекта преломления, чем наблюдалась в эксперименте.

Экспериментальные значения положения пичков представлены на Рис. 8 точками. Отметим, что эксперимент не предполагал исследование рассматриваемого эффекта, так что образец не был монокристаллическим вблизи поверхности. Наблюдаемая слабая зависимость положения пичков от резонансного преломления была случайным артефактом. Слабый эффект может быть объяснен тем, что пленка $YFe_2$ была закрыта слоем Nb, и при скользящих углах, когда излучение проникает на глубину ~ 3-5 нм, сигнал мог быть получен только из интерфейса $Nb/YFe_2$. Верхний слой Nb по свидетельству технологов имел мелкокристаллическую структуру, при этом интерфейс $Nb/YFe_2$ также мог быть достаточно разупорядочен. Это объясняет также и то, что ориентация самой пленки $YFe_2$ соответствовала <011>, при этом теоретически плоскости (111) должны были иметь угол с поверхностью ~ $35.3^{o}$, но в этом случае при скользящих углах мы бы не имели брэггговского отражения для межплоскостного расстояния d=0.4255 нм. Так что случайно зафиксированные брэгговские отражения могут быть объяснены как искажениями элементарной ячейки $YFe_2$ при внедрении атомов Nb, так и поликристаллизацией структуры $YFe_2$ в интерфейсной области. Восприимчивость в интерфейсной области также должна иметь ослабленную резонансную зависимость. Для такой несовершенной структуры нельзя было ожидать существенной резонансной зависимости положения брэгговских пиков от энергии фотонов, следующей из теории. Теоретические расчеты демонстрируют, что для экспериментов с идеальным кристаллом эта зависимость действительно может быть использована для восстановления магнитооптических констант.

Наибольший интерес в современных синхротронных исследованиях представляют магнитные эффекты в рассеянии, наблюдаемые вблизи краев поглощения. С учетом магнитного рассеяния рентгеновского излучения восприимчивость среды становится тензором, при этом определение недиагональных компонент этого тензора, отвечающих за магнитные эффекты в рассеянии наибольшее важно. Теория отражения и дифракции в общем случае существенно усложняется, но в некоторых случаях, например, при скользящем падении и ориентации намагниченности в плоскости рассеяния (L-MOKE геометрия), для волн круговой поляризации оказывается возможным использовать «скалярную» теорию отражения [31]. При этом в расчетах следует использовать восприимчивость с «магнитными добавками» $\chi \pm \Delta\chi_{magn}$, где знаки $\pm$ соответствуют правой и левой круговой поляризации падающего излучения. Влияние «магнитного» преломления на положение дифракционных максимумов проанализировано для значений $\Delta\chi_{magn}(E_{ph})$ из работы [15]. Рассчитанная зависимость разницы в положении брэгговского пика для двух круговых поляризаций $[\theta_m(E^+_{ph}) - \theta_m(E^-_{ph})]$ воспроизводит в общих чертах (обращенную) форму реальной части $\Delta\chi_{magn}(E_{ph})$ (Рис. 9). Результат, представленный на Рис. 9, демонстрирует, что зависимость $[\theta_m(E^+_{ph}) - \theta_m(E^-_{ph})]$ вполне наблюдаема даже для таких небольших значений наведенных магнитных моментов, как это имеет место на атомах Y в структуре $YFe_2$.

## 5. ЗАКЛЮЧЕНИЕ

Проведенный анализ показывает, что для идеальной структуры эффект преломления на положение дифракционного максимума в условиях резко асимметричной дифракции весьма значителен. Расчет положения дифракционного пика для правой и левой круговой поляризации с учетом магнитной добавки к восприимчивости из [15] показывает, что и магнитный эффект в условиях резко асимметричной дифракции для жесткого рентгеновского излучения наблюдаем.

Таким образом, резко ассиметричная дифракция может быть хорошим методом для определения магнитооптических констант в жестком диапазоне длин волн рентгеновского излучения.

Работа поддержана грантами РФФИ №№ 09-02-01293 и 13-02-00760-а
15-02-01502

.

## ЛИТЕРАТУРА

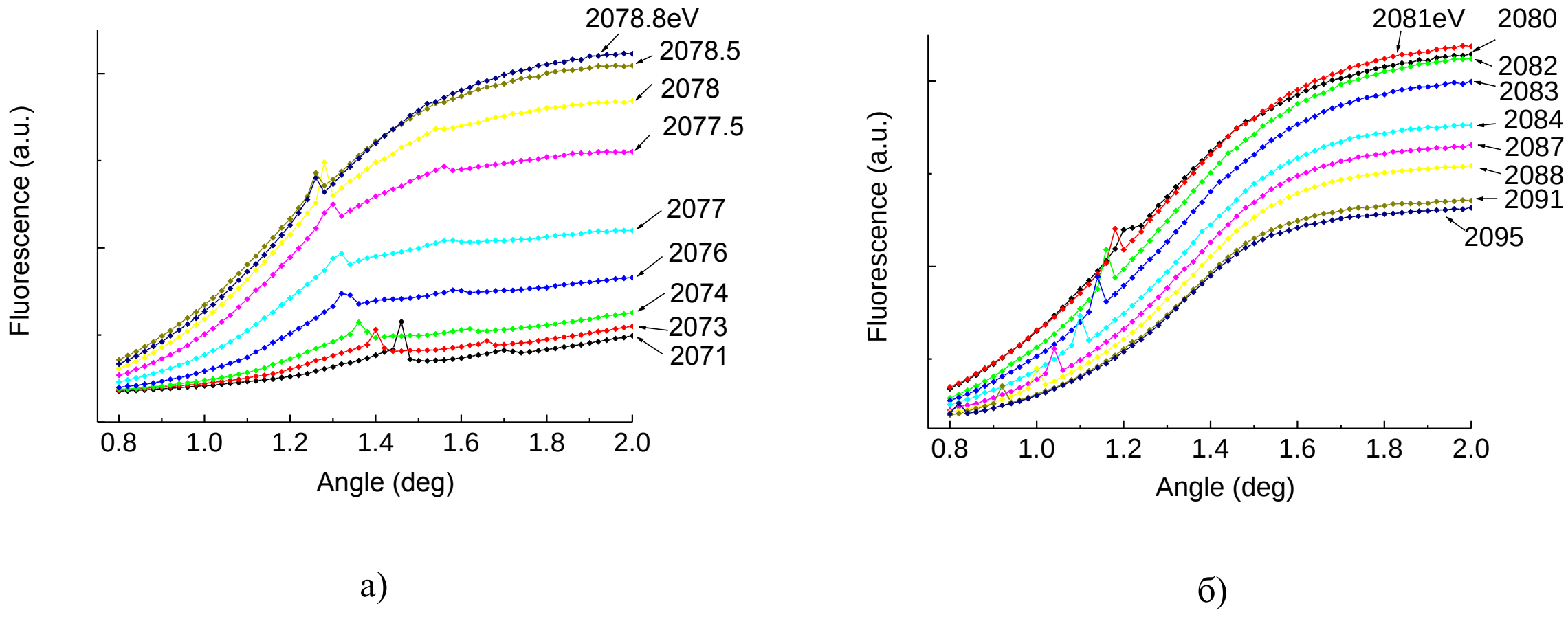

Рис. 1

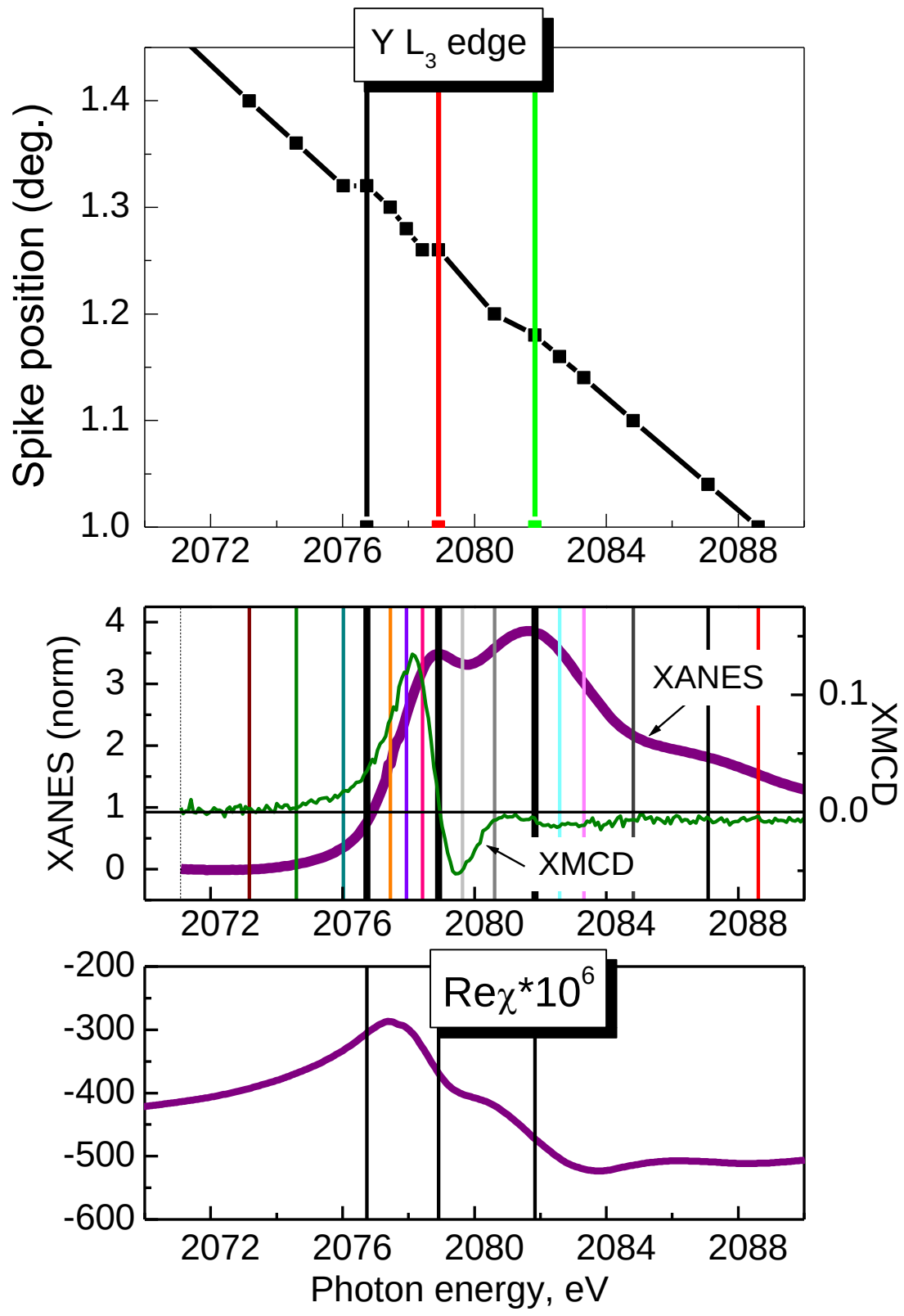

Рис. 2

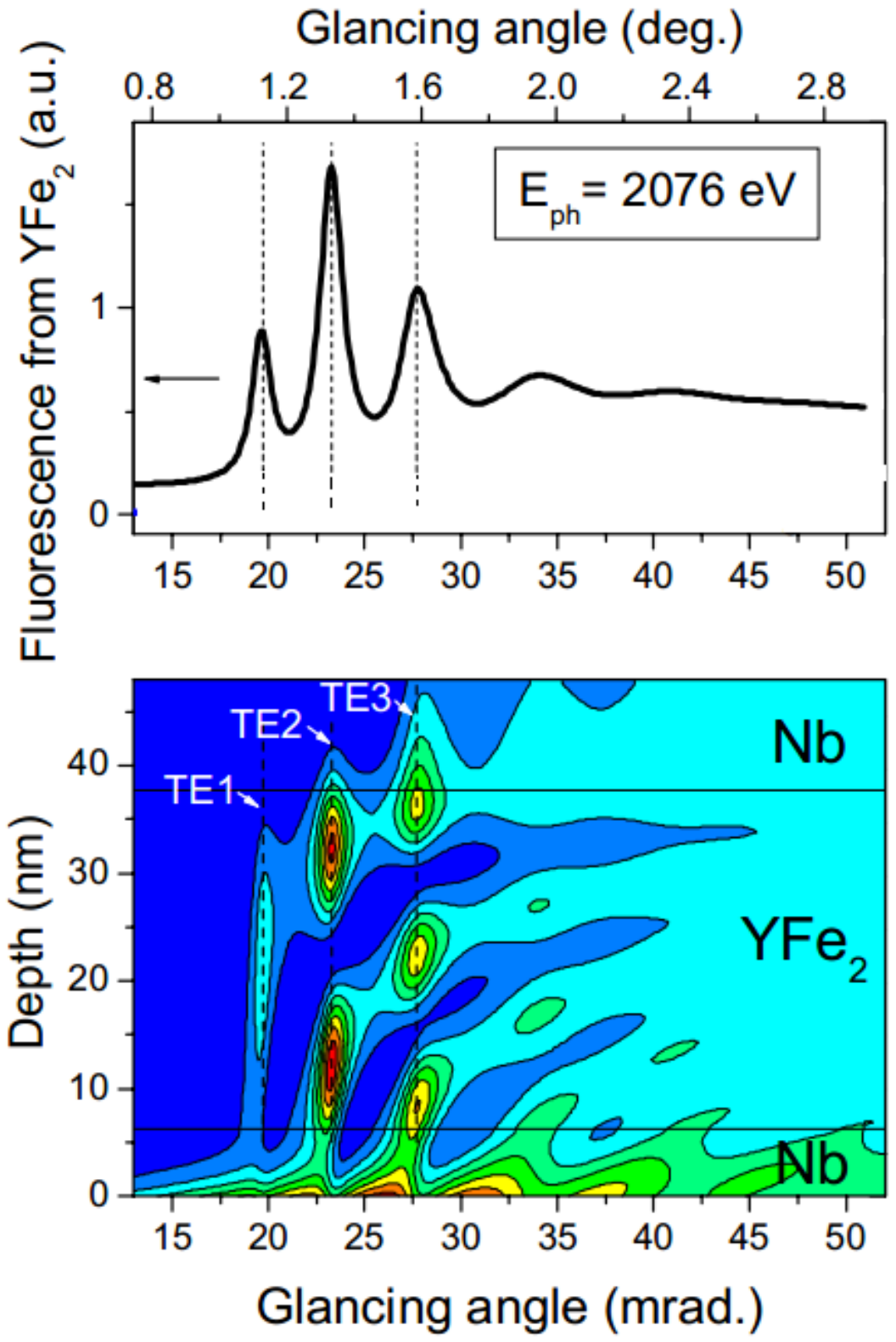

Рис. 3

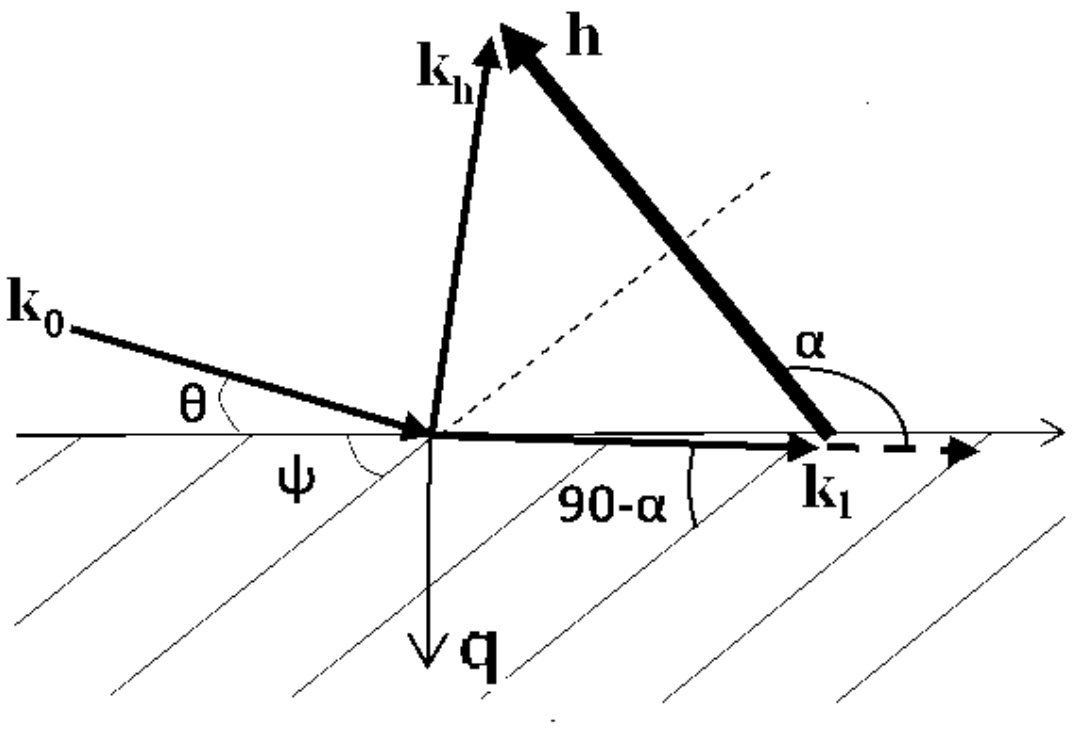

Рис. 4

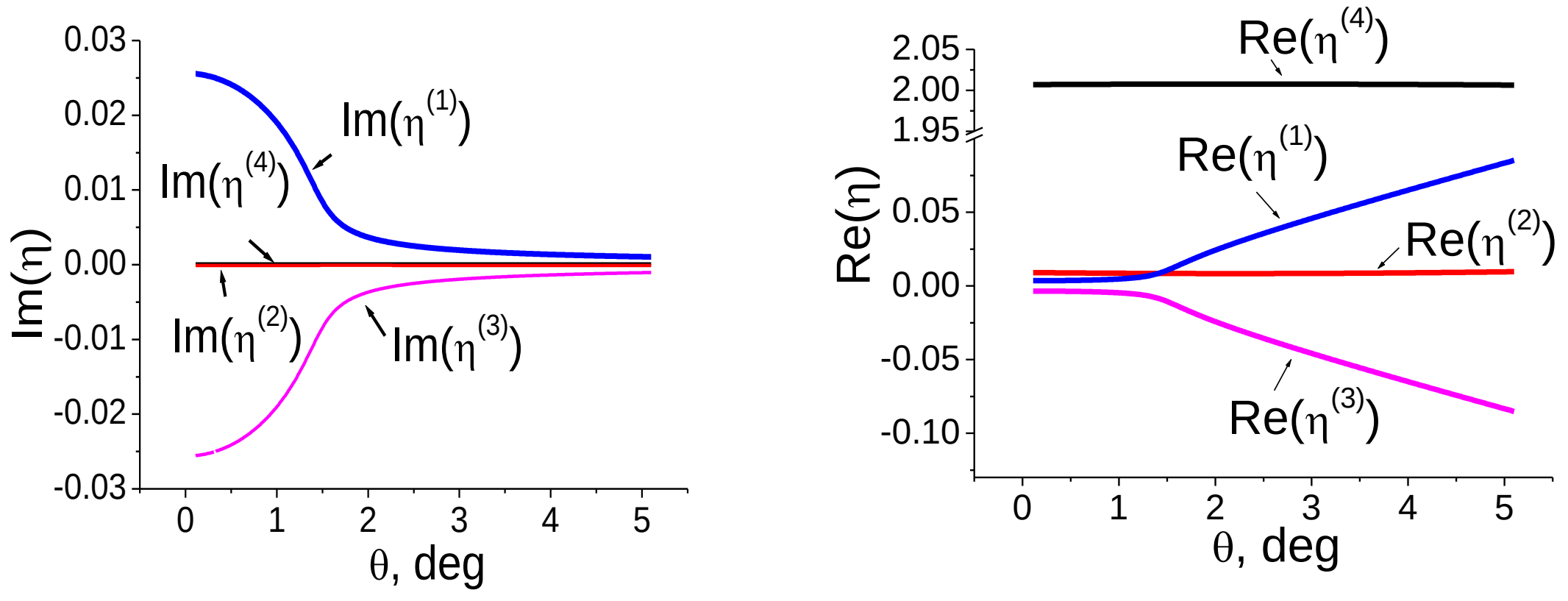

а) б)

Рис 5.

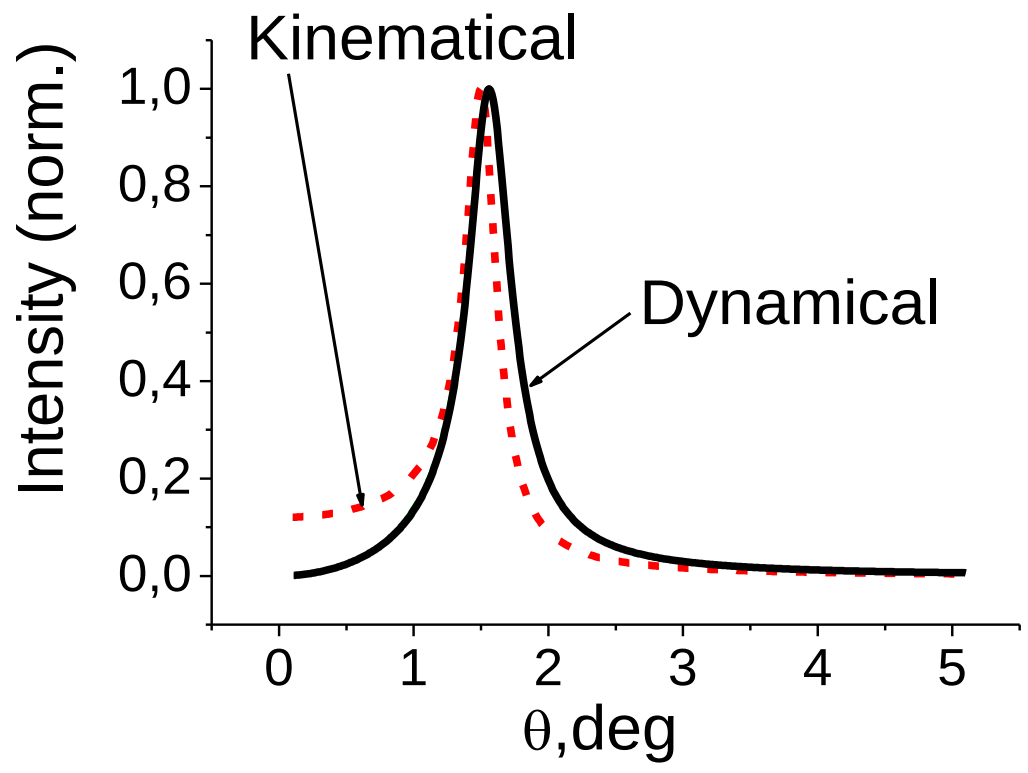

Рис 6.

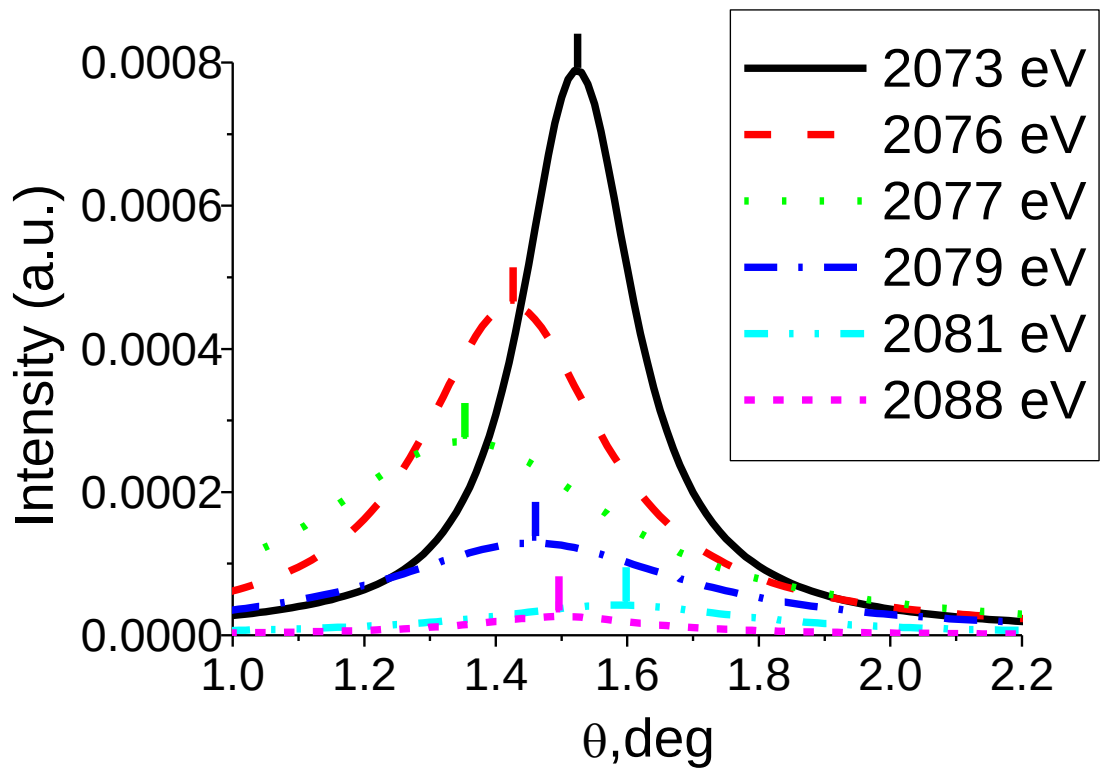

Рис. 7.

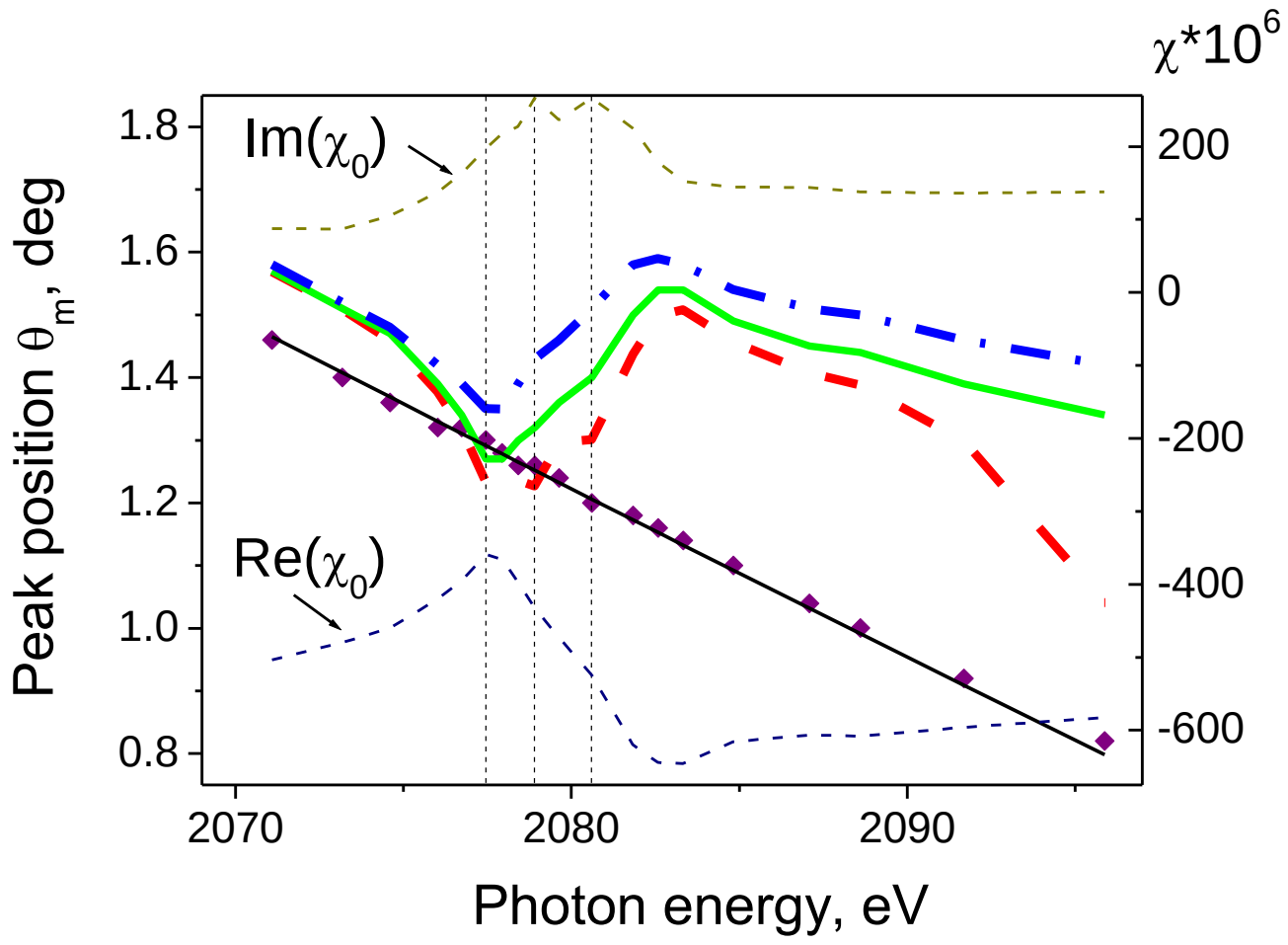

Рис. 8.

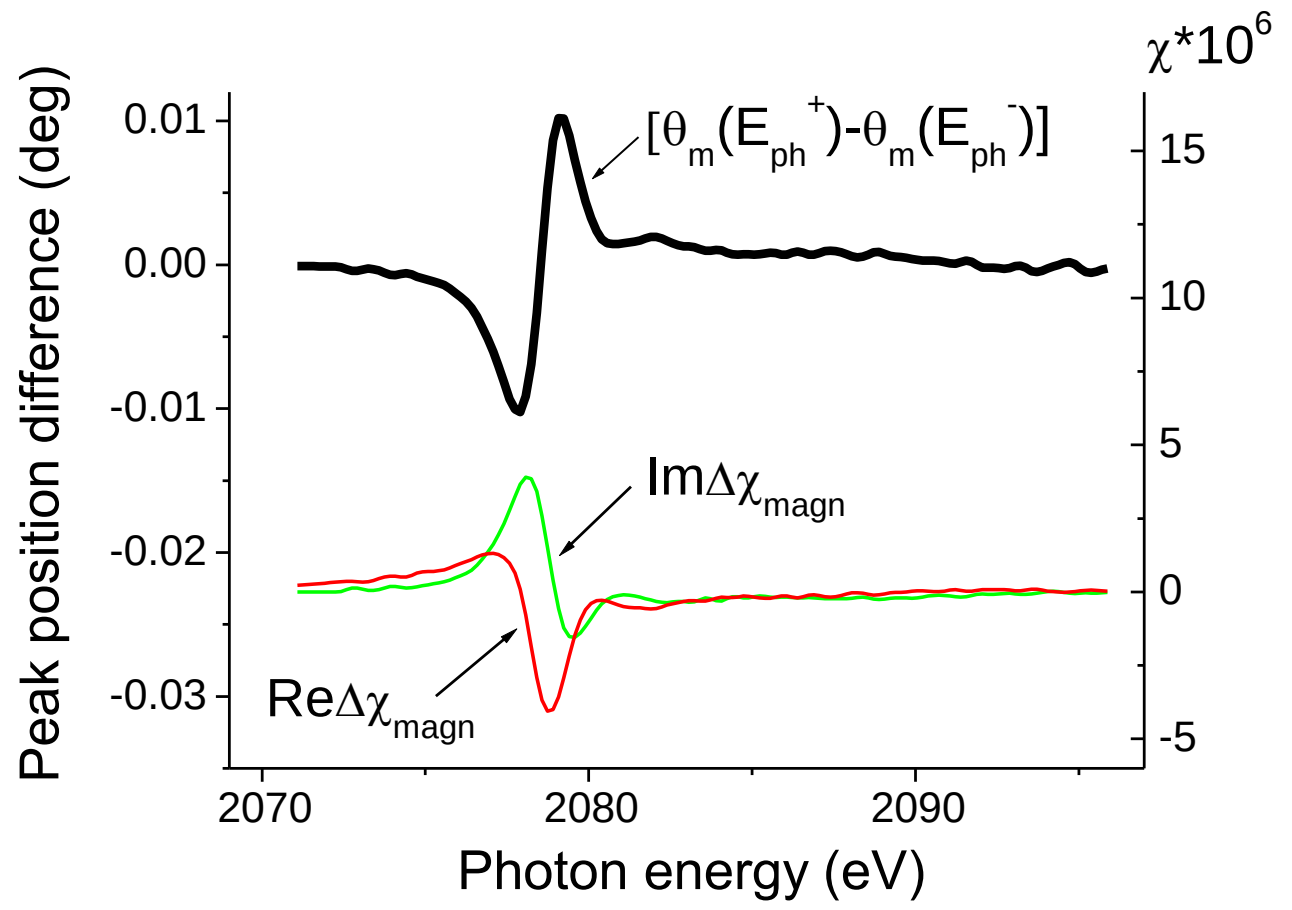

Рис. 9.

Подписи к рисункам

Рис. 1 (а,б). Экспериментальные зависимости выхода флуоресцентного излучения от образца Nb(4 нм)/$YFe_2$(40 нм <110>)/Fe(1.5 нм)/Nb(50 нм)/сапфир в функции угла скольжения падающего излучения $\theta$, измеренные для набора энергий фотонов вблизи $L_3$ края поглощения Y.

Рис. 2. Изменения угла скольжения $\theta_m$, при котором наблюдался пичок на кривых флуоресцентного выхода, в функции энергии падающих фотонов (верхний график, вертикальные линии обозначают некоторые резонансные особенности), экспериментальные зависимости поглощения и кругового дихроизма (XAS и XMCD) для нашего образца (средний график, вертикальные линии обозначают энергии падающих фотонов, для которых измерялись рефлектометрические и одновременно флуоресцентные кривые), зависимость реальной части восприимчивости, полученная при подгонке рефлектометрических кривых в работе [15] (нижний график).

Рис. 3. Попытка объяснить возникновение пичков на кривых выхода флуоресцентного излучения (верхний график) формированием волноводной моды в пленке $YFe_2$. На нижнем графике представлен квадрат модуля амплитуды поля рентгеновской стоячей волны в структуре Nb/$YFe_2$/Nb как функция угла скольжения и глубины – наибольшая интенсивность поля соответствует наиболее темным (красным в цветном варианте) областям.

Рис. 4. Рассматриваемая геометрия резко-асимметричной дифракции.

Рис. 5. Корни уравнения (13) для энергии фотонов $E_{ph}$=2082 эВ.

Рис. 6. Кривые дифракционного отражения (111) (в функции угла скольжения $\theta$) от кристалла $YFe_2$ для $E_{ph}$ =2082 eV, рассчитанные по формуле (8) и по динамической теории (13)-(17) в условиях резко асимметричной дифракции для $\psi = 43.9^o$.

Рис. 7. Кривые дифракционного отражения от кристалла $YFe_2$ для нескольких энергий фотонов вблизи $L_3$ края поглощения Y. Расчет по динамической теории. Вертикальные черточки отмечают положение максимумов, которое представлено на Рис. 8 в функции энергии фотонов.

Рис. 8. Расчет изменения угла скольжения $\theta_m$, при котором должен наблюдаться максимум брэгговского пика от идеального кристалла $YFe_2$, в функции энергии фотонов вблизи $L_3$ края поглощения Y: по обычному закону Вульфа-Брэгга (1) (тонкая черная линия), по обобщенной формуле (5) (пунктирная красная линия), с использованием функции Лауэ (8) (зеленая жирная линия) и по динамической теории (штрих-пунктирная синяя линия). Точками представлены экспериментальные значения положения пичков для образца Nb(4 нм)/$YFe_2$(40 нм <110>)/Fe(1.5 нм)/Nb(50 нм)/сапфир. Для сравнения на этом же рисунке приведены зависимости $\mathrm{Re}(\chi_0)$, $\mathrm{Im}(\chi_0)$ (правая шкала).

Рис. 9. Разница положения брэгговского пика для правой и левой круговой поляризации падающего излучения, рассчитанная для кристалла $YFe_2$ вблизи $L_3$ края поглощения Y. Для сравнения на этом же графике приведены спектральные зависимости $\Delta\chi_{magn}(E_{ph})$ из работы [15] (правая шкала).

Автореферат

Дано объяснение наблюдавшихся в эксперименте пиков на кривых выхода флуоресценции в функции угла скольжения от образца Nb(4 нм)/$YFe_2$(40 нм <110>)/Fe(1.5 нм)/Nb(50 нм)/сапфир для падающего излучения с энергией фотонов в окрестности $L_3$ края поглощения иттрия. Изменение положения пика в функции энергии фотонов согласуется с предположением о том, что в детекторе одновременно с флуоресцентным сигналом регистрируется дифракционное отражение (111) от кристалла $YFe_2$. Рассмотрена теория дифракции в случае резко асимметричной дифракции в кинематическом и динамическом приближениях. Получено доказательство, что эффект преломления при скользящих углах падения существенно влияет на положение пика дифракционного отражения от кристалла в обоих приближениях, так что смещение этого пика в функции энергии фотонов может быть использовано для определения спектральной зависимости магнитооптических констант в окрестностях краев поглощения рентгеновского излучения.

Для переписки:

Андреева Марина Алексеевна
e-mail: Mandreeva1@yandex.ru

Адрес:
Москва 119991 Ленинские горы 1, стр.2,
Физический факультет МГУ имени М.В.Ломоносова
Кафедра физики твердого тела
Андреевой Марине Алексеевне

Тел.: 8 903 7120837 (моб), 8 495 8511157 (дом)